# Mechanical Properties of Formamidinium Halide Perovskites $FABX_3$ (FA = $CH(NH_2)_2$; B = Pb, Sn; X = Br, I) From First-Principles

Lei Guo[1], Gang Tang[1] and Jiawang Hong[1,*]

**Abstract** The mechanical properties of formamidinium halide perovskite $FABX_3$ (FA = $CH(NH_2)_2$; B = Pb, Sn; X = Br, I) were systematically investigated by using the first-principles calculations. Our results reveal that $FABX_3$ perovskites possess excellent mechanical flexibility, ductility and strong anisotropy. It shows that the planar organic cation $FA^+$ has an important effect on the mechanical properties of $FABX_3$ perovskites. In addition, our results indicate that: (i) the moduli (bulk modulus $B$, Young's modulus $E$, and shear modulus $G$) of $FABBr_3$ are larger than those of $FABI_3$ for the same B atom and (ii) the moduli of $FAPbX_3$ are larger than those of $FASnX_3$ for the same halide atom. The reason of the two trends was demonstrated by carefully analyzing the bond strength between B and X atom based on the projected crystal orbital Hamilton population method.

During the last several years, the power conversion efficiency of organic-inorganic hybrid perovskites (OIHPs) ABX$_3$ (A = MA (CH$_3$NH$_3$), FA (CH(NH$_2$)$_2$); B = Pb, Sn; X = I, Br) has increased from 3.81% to 23.7%,[1-5] which attracts tremendous attentions on the understanding of superior photovoltaic performances and various potential optoelectronic applications. Some experimental and theoretical investigations have confirmed that the excellent optoelectronic properties of OIHPs result from several advantages, such as the suitable direct band gap, defect-tolerance, long carrier lifetime and diffusion length etc.[6-8] In addition to the advantages mentioned above, another advantage of OIHPs compared to their inorganic counterparts is the soft nature of the hybrid framework, providing the possible applications in flexible and wearable electronic devices. Recently, it was reported that stress and strain existing in hybrid perovskites play a critical role on film stability and device performance.[9, 10] Consequently, the researches on the mechanical property and deformation behaviour of OIHPs are attracting widespread attention.

Feng and Roknuzzaman et al. theoretically investigated the mechanical properties of methylammonium halide perovskite MABX$_3$ (B = Sn, Pb; X = Br, I) by using the first-principles method.[11, 12] Subsequently, some experimental techniques such as nanoindentation, sound velocity measurement, and inelastic neutron scattering, were also employed to investigate the elastic properties of organic-inorganic hybrid perovskites MABX$_3$( B = Pb; X = I, Br) at room temperature.[13-15] Compared to MABX$_3$, formamidunium halide perovskite FABX$_3$ have more optimal electronic band gaps, superior thermal stability, and a lower hysteresis.[16-18] However, the mechanical properties of FABX$_3$ need further investigations to understand its flexible nature.

In this work, the mechanical properties and bonding characteristics of FABX$_3$ (B = Sn, Pb; X = Br, I) compounds were systematically investigated by using the first-principles calculations. Our results show that FABX$_3$ perovskites are mechanically flexible and ductile. It is also observed that FABX$_3$ perovskites exhibit strong anisotropy on the mechanical properties. We find that the organic cation FA$^+$ structure plays an important role in the shear elastic constants. Finally, two general tendencies are obtained as follows: (i) replacing I with Br, the values of moduli (bulk modulus *B*, Young's modulus *E*, and shear modulus *G*) increase in Pb-based or Sn-based perovskites. (ii) replacing Pb with Sn, the values of moduli decrease in I-based or Br-based perovskites. The reasons were revealed by carefully analyzing the B-X bonding characteristics in FABX$_3$ based on the projected crystal orbital Hamilton population method.

Our calculations are performed by using the Vienna Ab initio Simulation Package (VASP) code in the framework of density function theory (DFT).[19, 20] The electron-ion interaction is described by the projector augmented wave (PAW) method.[21] The plane-wave cut-off energy was set to 840 eV. The 9×9×9 Monkhorst-Pack *k*-point mesh was employed for sampling the Brillouin zone. Both *k*-point mesh and the cut-off energy were tested carefully for the convergence. The lattice parameters and atomic positions were fully relaxed until the energy difference is less than $10^{-4}$ eV and the force on each atom is smaller than $10^{-2}$ eV/ Å. In this work, the PBE+vdW-DF2 method was used to consider the Van der Waals (vdW) interactions, which plays an important role in the hybrid perovskites materials with weak interactions along the stacking direction.[22, 23] We utilized the package Lobster to compute the crystal orbital Hamiltonian population (COHP) for the bonding analysis.[24, 25] Mechanical properties are calculated by finite difference method in VASP.[26]

According to previous experiments, FABX$_3$ (FA = CH(NH$_2$)$_2$; B = Pb, Sn; X = I, Br) mainly adopts cubic crystal structures at the room temperature.[27] Our calculations are based on the FAPbI$_3$ cubic structure obtained from experiment (space group: *Pm3m*; lattice constant: 6.36 Å; atomic position: Pb (0.0, 0.0, 0.0), I (0.5, 0.0, 0.0), C (0.5, 0.57, 0.5), N (0.68, 0.48, 0.5), H (0.5, 0.74, 0.5), H (0.81, 0.57, 0.5), H (0.70, 0.32, 0.5)).[27] The crystal structure of FABX$_3$ is shown in Figure 1. The center of octahedron is occupied by B atom, the X atoms share the corner positions of octahedron, and the organic cation FA lies in a cage surrounded by [PbX$_6$] octahedrons. Following previous theoretical work,[28, 29] we fully relaxed the atomic coordinates and lattice parameters from the experimental cubic structure and obtained the pseudo-cubic structure. As can be seen in Table I ($X_a$, $X_b$, $X_c$ represents the halide X atom along *a*, *b*, *c* axis, respectively), the calculated lattice constants are slightly larger than experimental values[27, 30-32] owing to the overestimation of the PBE method. We can also see that replacing I (Pb) with smaller atoms Br (Sn), the lattice constant of $FABX_3$ decreases, showing the same trend as the experimental values. Due to the influence of organic cation $FA^+$, the B-$X_{a(c)}$-B angles are not equal to 180°, illustrating that $X_{a(c)}$ is away from *a (c)* axis direction and lead to slight distortion of the $FABX_3$ structure.

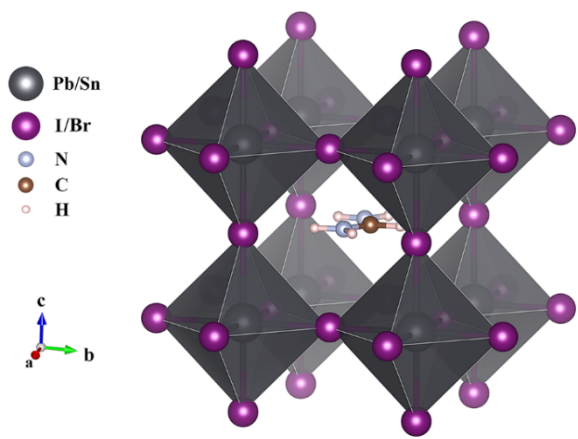

Fig. 1. The cubic structure of $FABX_3$ (FA = $CH(NH_2)_2$; B = Pb, Sn; X = I, Br).

Table I. Lattice parameters and atomic bonds of $FABX_3$ (FA = $CH(NH_2)_2$; B = Pb, Sn; X = Br, I) perovskites.

| | FABX₃ | | FAPbI₃ | FAPbBr₃ | FASnI₃ | FASnBr₃ |
|---|---|---|---|---|---|---|
| Lattice constant (Å) | This work | *a* | 6.67 | 6.30 | 6.62 | 6.26 |
| | | *b* | 6.50 | 6.08 | 6.46 | 6.03 |
| | | *c* | 6.61 | 6.21 | 6.53 | 6.14 |
| | Expt.[27, 30-32] | a/b/c | 6.36 | 5.99 | 6.32 | 6.03 |
| Bond angle (degree) | | B-$X_a$-B | 178.9 | 178.0 | 176.7 | 179.0 |
| | | B-$X_b$-B | 180.0 | 180.0 | 180.0 | 180.0 |
| | | B-$X_c$-B | 174.4 | 175.0 | 171.5 | 172.5 |
| Bond length (Å) | | B-$X_a$ | 3.34 | 3.15 | 3.31 | 3.13 |
| | | B-$X_b$ | 3.22 | 3.03 | 3.04 | 2.83 |
| | | B-$X_c$ | 3.31 | 3.11 | 3.28 | 3.08 |

The elastic constants of $FABX_3$ perovskites were obtained from the first-principles calculations, as shown in Table II. Our calculated elastic constants of these hybrid halide perovskites satisfy the elastic stability conditions, suggesting all structure are mechanically stable.[33-35] The calculated $C_{11}$, $C_{44}$ and $C_{12}$ by this work are slightly larger than experimental results[13] and data calculated by Wang et al..[36] $C_{11}$ ($C_{22}$ or $C_{33}$) are rather larger than other elastic constants, implying strong resistances to stretch along [100] ([010] or [001]) direction but weak resistances to shear deformation for $FABX_3$ compounds. This is consistent with the observation from recent experimental works.[13] Interestingly, we note that $C_{55}$ is relatively larger than $C_{44}$ and $C_{66}$ for all compounds which implies a better ability to resist shear deformation in (010) crystal plane than (100) and (001) planes. $C_{13}$ is also relatively larger compared to $C_{12}$ and $C_{23}$. The larger $C_{55}$ ($C_{13}$) than $C_{44}$, $C_{66}$ ($C_{12}$, $C_{23}$) indicates the stronger couplings between $a$, $c$ crystal directions than those between $a$, $b$ and $b$, $c$ directions. That may result from the halide X which distort away from $a$ and $c$ crystal axis in such pseudo-cubic structure due to the orientation of planar organic cation $FA^+$.

The elastic moduli are calculated by following formulas:[35]

$$B_V = \frac{1}{9}(C_{11}+C_{22}+C_{33}) + \frac{2}{9}(C_{12}+C_{13}+C_{23}), \quad (1)$$

$$G_V = \frac{1}{15}(C_{11}+C_{22}+C_{33}) - \frac{1}{15}(C_{12}+C_{13}+C_{23}) + \frac{1}{5}(C_{44}+C_{55}+C_{66}), \quad (2)$$

$$B_R = \frac{1}{(S_{11}+S_{22}+S_{33})+2(S_{12}+S_{13}+S_{23})}, \quad (3)$$

$$G_R = \frac{15}{4(S_{11}+S_{22}+S_{33})-4(S_{12}+S_{13}+S_{23})+3(S_{44}+S_{55}+S_{66})}, \quad (4)$$

where the $S_{ij}$ is the elastic compliance matrix. The bulk modulus ($B$), shear modulus ($G$), Young's modulus ($E$), Poisson's ratio($v$), Pugh's ratio ($B/G$)[37], and the universal anisotropy index $A^U$ [38] of these perovskite compounds were calculated using the Voigt-Reuss-Hill (VRH) approximation, the relationships are given as:

$$B = \frac{B_V+B_R}{2}, \quad (5)$$

$$G = \frac{G_V+G_R}{2}, \quad (6)$$

$$E = \frac{9BG}{(3B+G)}, \tag{7}$$

$$v = \frac{(3B-2G)}{[2(3B+G)]}, \tag{8}$$

$$A^U = 5\frac{G_V}{G_R} + \frac{B_V}{B_R} - 6, \tag{9}$$

As shown in Table III, the calculated bulk modulus of $FAPbBr_3$ (18.34 GPa) is in good agreement with experimental value (16.9±1.7 GPa).[13] In terms of Young's modulus $E$, our results are relatively larger than experimental values.[15] The ductility of a material has been extensively indexed by the critical value 1.75 for Pugh's ratio (0.26 for Poisson's ratios). In other words, the material is ductile (brittle) if Pugh's ratio or Poisson's ratios is larger (smaller) than 1.75 or 0.26. [11], [37] Our results show that all the calculated perovskites have large $B/G$ (from 2.37 to 2.88) and $v$ (from 0.31 to 0.34), which indicates that $FABX_3$ exhibit a good ductility suitable for flexible materials. Moreover, the Br-based perovskites in this study ($FAPbBr_3$ and $FASnBr_3$) have largest Pugh's ratio (larger than 2.8), suggesting they are the most ductile materials in these compounds. In addition, we notice that the Br-based perovskites exhibit more strong anisotropic properties compared with I-based perovskites according to the universal anisotropy index $A^U$, as well as the three dimensional (3D) surface contours of Young's modulus (Figure 2). This strong anisotropy indicates Br-based perovskites will be more unstable than I-based perovskites at ambient conditions.[11, 39] Compared with MA-based organic-inorganic perovskites,[11] all of them possess excellent mechanical flexibility, ductility and strong anisotropy. In addition, the Young's moduli of FA-based perovskites are lower than that of MA-based perovskites, which is due to larger $FA^+$ weakening the inorganic framework.[15]

Table II. Elastic constants of $FABX_3$ perovskite compounds from calculations and experiment. the unit is GPa.

| $FABX_3$ | $C_{11}$ | $C_{22}$ | $C_{33}$ | $C_{44}$ | $C_{55}$ | $C_{66}$ | $C_{12}$ | $C_{13}$ | $C_{23}$ |
|---|---|---|---|---|---|---|---|---|---|
| $FAPbI_3$ | 30.15 | 31.00 | 29.85 | 2.03 | 5.33 | 2.60 | 2.99 | 7.22 | 4.26 |
| Cal.[36] | 20.50 | | | 4.80 | | | 12.30 | | |

| FABX$_3$ | | | | | | | | | |
|---|---|---|---|---|---|---|---|---|---|
| FAPbBr$_3$ | 37.98 | 45.21 | 34.59 | 3.34 | 5.30 | 1.85 | 7.26 | 9.95 | 5.65 |
| Expt.[13] | 31.2±0.2 | | | 1.5±0.1 | | | 4± 0.5 | | |
| Expt.[13] | 27.7±1.6 | | | 3.1±0.1 | | | 11.5±2.4 | | |
| FASnI$_3$ | 29.96 | 25.21 | 26.02 | 2.35 | 4.90 | 2.41 | 6.85 | 8.23 | 3.28 |
| FASnBr$_3$ | 35.34 | 26.66 | 32.13 | 3.29 | 5.65 | 1.21 | 5.72 | 10.85 | 3.77 |

Table III. Calculated and experimental mechanical parameters of FABX$_3$, including Bulk modulus (*B*), shear modulus (*G*), Young's modulus (*E*), Poisson's ratio (*v*), Pugh's ratio (*B/G*), the universal anisotropy index $A^U$. For the mechanical moduli, the unit is GPa.

| FABX$_3$ | *B* | *G* | *E* | *B/G* | *v* | $A^U$ |
|---|---|---|---|---|---|---|
| FAPbI$_3$ | 13.25 | 5.59 | 14.70 | 2.37 | 0.31 | 3.68 |
| Cal.[36] | 15.30 | 3.60 | 9.90 | 4.30 | 0.4 | |
| Expt. | | | (11.8±1.9)[15] | | | |
| FAPbBr$_3$ | 18.34 | 6.37 | 17.11 | 2.88 | 0.34 | 4.75 |
| Expt. | (16.9±1.7)[13] | | (12.3±0.8)[15] | | | |
| FASnI$_3$ | 12.96 | 5.08 | 13.48 | 2.55 | 0.33 | 2.60 |
| FASnBr$_3$ | 14.64 | 5.18 | 13.88 | 2.83 | 0.34 | 5.27 |

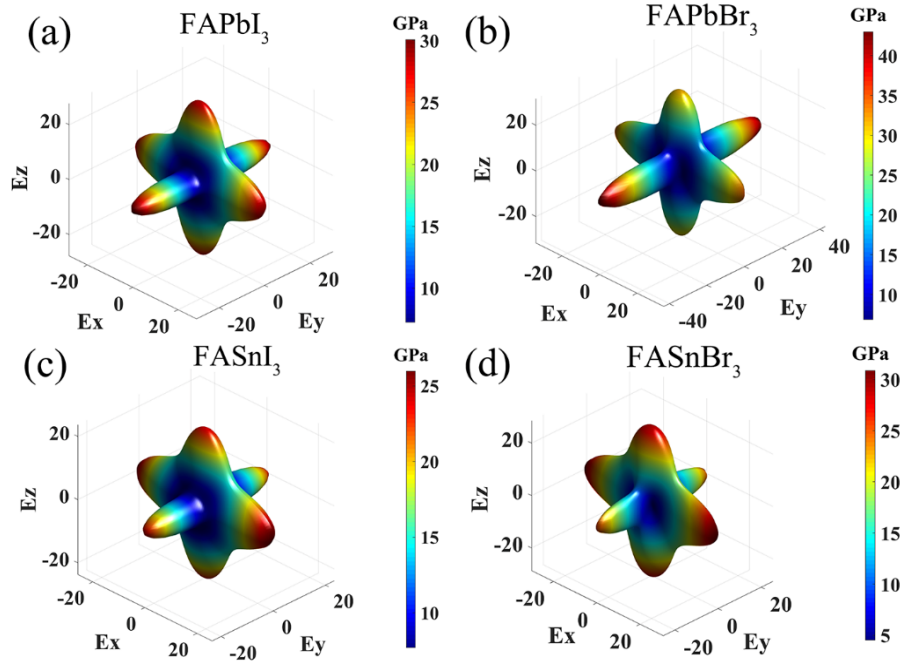

Fig. 2. Three dimensional (3D) surface contour of Young's modulus for (a) FAPbI$_3$, (b) FAPbBr$_3$,

(c) FASnI$_3$, (d) FASnBr$_3$.

We also observe two general tendencies: (i) replacing I with Br, the values of bulk modulus, shear modulus and Young's modulus increase in Pb/Sn-based perovskites. (ii) replacing the Pb with Sn, the values of bulk modulus, shear modulus and Young's modulus decrease in I/Br-based perovskites. The first tendency is in consistence with recent experimental results.[15] For the second tendency, there are no experimental reports about mechanical properties of Sn-based perovskites. However, we noticed that it was reported from recent first-principles calculations that the bulk modulus and shear modulus of Pb-based perovskites are also larger than Sn-based perovskites in MABX$_3$ or CsBX$_3$ compounds, which is in line with our results.[12, 40]

The first tendency can be explained in terms of the inorganic B-X bonds in FABX$_3$ perovskites, which is listed in Table I. We can see that the B-X bond length in Br-based perovskites is shorter than that in I-based perovskites, which means that B-Br bond is stronger than B-I bond due to the smaller radius of Br compared with I. However, this bond-length analysis can't explain the second tendency. As can be seen from Table I to III, the Sn-based perovskites with shorter bond lengths than those in Pb-based compounds have even smaller elastic moduli. Though Sn has smaller radius than Pb atom, its electronegativity[11], [41] is also smaller (Figure 3), this will induce the weaker interactions between Sn-X than Pb-X. Therefore, Sn-based perovskites show smaller elastic moduli than Pb-based materials. The bond interactions can be more accurately calculated from the projected crystal orbital Hamilton population (pCOHP) approach[42, 43], which will be discussed in more detailed in next section.

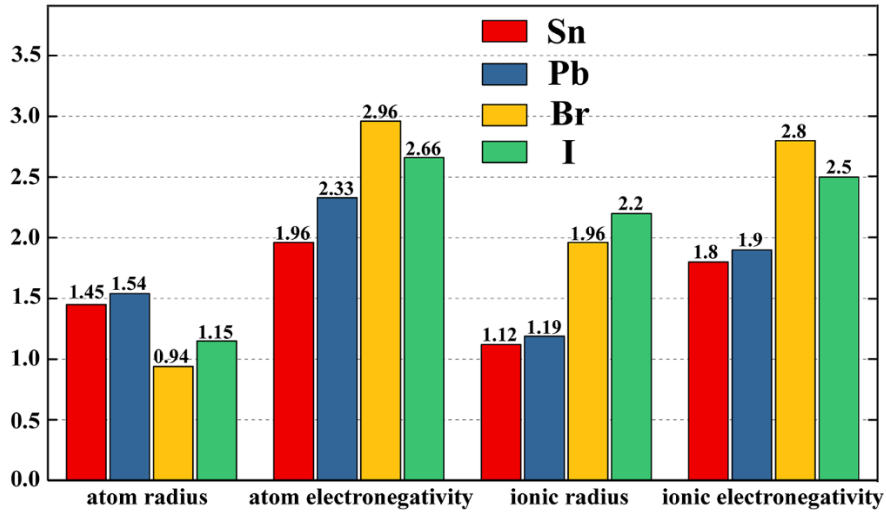

Fig. 3.   The radius (Å) and electronegativity of Sn, Pb, Br and I atoms and ions.

To accurately understand the bonding characteristics of FAB$X_3$ perovskites, such as B-X bonding, the newly developed projected crystal orbital Hamilton population (pCOHP) approach[42, 43] is employed, which was recently introduced to investigate the hydrogen bonding in the hybrid perovskite.[39, 44] Negative pCOHP values indicate bonding states, positive pCOHP values indicate anti-bonding states. The integrated COHP (ICOHP) which reflects the bond strength (the more negative value, the stronger the bonding strength) was also calculated.[44]

The pCOHP and ICOHP averaged over three B-X (B-$X_a$, B-$X_b$, B-$X_c$) atom pairs in FAB$X_3$ compounds were shown in Figure 4. The ICOHP value of Pb-Br bonding is -2.356 eV, which is stronger than Pb-I bonding (-2.115 eV) in Pb-based perovskites, and the ICOHP value of Sn-Br (-2.237 eV) is also more negative than that of Sn-I (-2.056 eV), which provide a good explanation for the first tendency of moduli concluded above. As for replacing Pb with Sn in the compositions that have same halogen atoms, it is found that ICOHP value of Pb-I (-2.115 eV) is more negative than that of Sn-I (-2.056 eV) in I-based perovskites, and ICOHP value of Pb-Br (-2.356 eV) is also more negative than that of Sn-Br (-2.237 eV) in Br-based perovskites, illustrating that Pb-based perovskites exhibit stronger B-X bond compared with Sn-based perovskites, which well explains the second tendency in previous section.

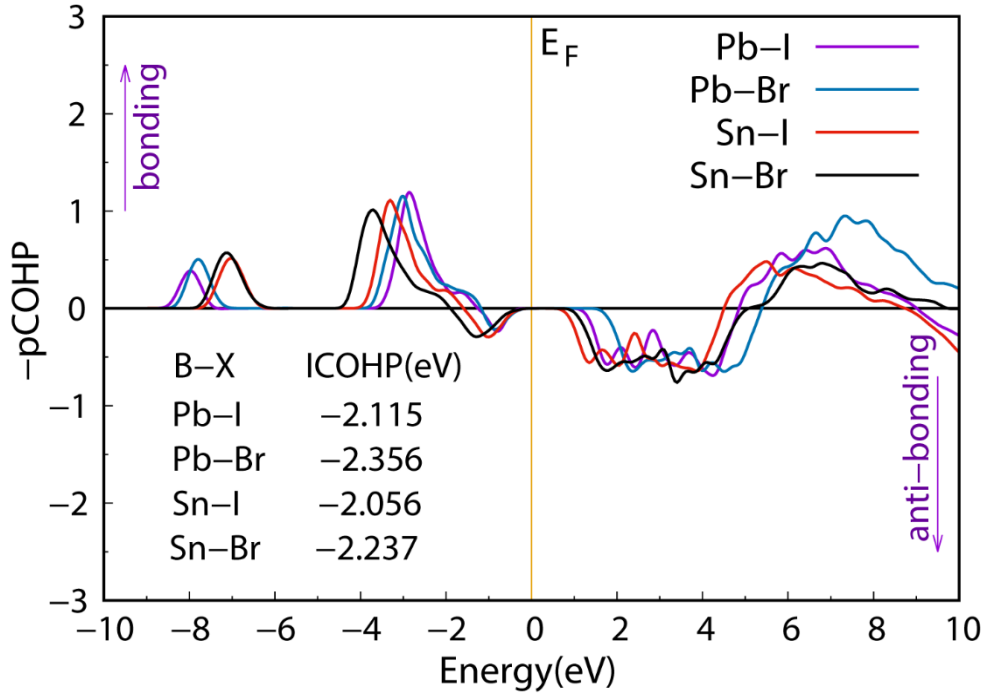

Fig. 4. Crystal orbital Hamilton population (COHP) and integrated COHP (ICOHP) analysis of FABX$_3$ (FA = CH(NH$_2$)$_2$; B = Pb, Sn; X = I, Br).

In summary, the mechanical properties of FABX$_3$ (FA = CH(NH$_2$)$_2$; B = Pb, Sn; X = Br, I) were investigated using first-principles calculations. Our calculation results reveal that these hybrid perovskite materials are flexible, ductile, and they are suitable for compliant devices with large deformation demanded. The universal anisotropy index $A^U$ demonstrates that these FABX$_3$ perovskites have a very strong anisotropy. And it is found that the orientation of the planar organic cation FA$^+$ has an effect on the structure and mechanical properties of FABX$_3$ perovskites. Further, our results indicate that: (i) replacing the I with Br, the values of bulk modulus, shear modulus and Young's modulus increase in Pb-based or in Sn-based perovskites. (ii) replacing the Pb with Sn, the values of bulk modulus, shear modulus and Young's modulus decrease in I-based or Br-based perovskites. The reasons are revealed by analyzing bonding characteristics including the atomic radius, electronegativity, and projected crystal orbital Hamilton population of FABX$_3$ perovskite compounds. The case of Pb-based and Sn-based materials show that it is not enough to analyze the mechanical properties from sole bonding length information, more bonding characteristics, such as electronegativity and crystal orbital Hamilton population are needed to be taken into account. Our work may shed lights on the future experimental works on mechanical properties of hybrid halide perovskites and their photovoltaic applications.

**AUTHOR INFORMATION**

**Corresponding Author**

*E-mail: hongjw@bit.edu.cn.

**Acknowledgements**

This work is supported by the National Science Foundation of China (Grant No. 11572040) and the Thousand Young Talents Program of China. Theoretical calculations were performed using resources of National Supercomputer Centre in Guangzhou.

**References**

[1] Julian B, Norman P, Soo-Jin M et al 2013 *Nature* **499** 316

[2] Heo J H, Im S H, Noh J H et al 2013 *Nat. Photonics* **7** 486

[3] https://www.nrel.gov/pv/assets/pdfs/pv-efficiency-chart.20181221.pdf

[4] Kojima A, Teshima K, Shirai Y et al 2009 *J. Am. Chem. Soc.* **131** 6050

[5] Yang W S, Park B W, Jung E H et al 2017 *Science* **356** 1376

[6] Liu M Z, Johnston M B, Snaith H J 2013 *Nature* **501** 395

[7] Shi D, Adinolfi V, Comin R et al 2015 *Science* **347** 519

[8] Jiang M, Wang L, Xie H M Qiu Y Q 2018 *Chin. Phys. B* **27** 67102

[9] Rolston N, Bush K A, Printz A D et al 2018 *Adv. Energy Mater.* **8** 1802139

[10] Zhao J J, Deng Y H, Wei H T et al 2017 *Sci Adv.* **3** eaao5616

[11] Feng J 2014 *APL Mater.* **2** 081801

[12] Roknuzzaman M, Ostrikov K, Wasalathilake K C et al 2018 *Org. Electron.* **59** 99

[13] Ferreira A C, Letoublon A, Paofai S et al 2018 *Phys. Rev. Lett.* **121** 085502

[14] Sun S J, Fang Y N, Kieslich G et al 2015 *J. Mater. Chem. A* **3** 18450

[15] Sun S J, Isikgor F H, Deng Z Y et al 2017 *ChemSusChem* **10** 3740

[16] Eperon G E, Bryant D, Troughton J et al 2015 *J. Phys. Chem. Lett* **6** 129

[17] Eperon G E, Stranks S D, Menelaou C et al 2014 *Energy Environ. Sci.* **7** 982

[18] Zhang Y Y, Chen S Y, Xu P et al 2018 *Chin. Phys. Lett.* **35** 036104

[19]   Kohn W and Sham L J 1965 *Phys. Rev.* **140** A1133

[20]   Kresse G and Furthmüller J 1996 *Comput. Mater. Sci.* **6** 15

[21]   Blöchl P E 1994 *Phys. Rev. B* **50** 17953

[22]   Ahmad S, Kanaujia P K, Niu W et al 2014 *ACS Appl. Mater. Interfaces* **6** 10238

[23]   Lee K, Murray E D, Kong L et al 2010 *Phys. Rev. B* **82** 081101

[24]   Deringer V L, Tchougréeff A, Dronskowski R 2011 *J. Phys. Chem. A* **115** 5461

[25]   Maintz S, Deringer V L, Tchougréeff A L et al 2016 *J. Comput. Chem.* **37** 1030

[26]   Yvon L P, Paul S 2002 *Phys. Rev. B* **65** 104104

[27]   Weller M T, Weber O J, Frost J M et al 2015 *J. Phys. Chem. Lett.* **6** 3209

[28]   Geng W, Zhang L, Zhang Y N et al 2014 *J PHYS CHEM C* **118** 19565

[29]   Jong U G, Yu C J, Ri G C et al 2018 *J. Mater. Chem. A* **6** 1067

[30]   Hanusch F C, Wiesenmayer E, Mankel E et al 2014 *J. Phys. Chem. Lett* **5** 2791

[31]   Mitzi D B, Liang K 1997 *J. Solid State Chem.* **134** 376

[32]   Ferrara C, Patrini M, Pisanu A et al 2017 *J. Mater. Chem. A* **5** 9391

[33]   Ding Y C, Chen M, Gao X Y et al 2012 *Chin. Phys. B* **21** 067101

[34]   Wu Z J, Zhao E J, Xiang H P et al 2007 *Phys. Rev. B* **76** 054115

[35]   Ravindran P, Fast L, Korzhavyi P A et al 1998 *J. Appl. Phys.* **84** 4891

[36]   Wang J F, Fu X N, Wang J T 2017 *Chin. Phys. B* **26** 106301

[37]   Pugh S F 1954 *The London, Edinburgh, and Dublin Philosophical Magazine and Journal of Science* **45** 823

[38]   Ranganathan S I, Ostoja-Starzewski M 2008 *Phys. Rev. Lett.* **101** 055504

[39]   Tang G, Yang C, Stroppa A et al 2017 *J. Chem. Phys.* **146** 224702

[40]   Roknuzzaman M, Ostrikov K, Wang H et al 2017 *Sci. Rep.* **7** 14025

[41]   Allred A L 1961 *J. Inorg. Nucl. Chem.* **17** 215

[42]   Amat A, Mosconi E, Ronca E et al 2014 *Nano Lett.* **14** 3608

[43]   Kato M, Fujiseki T, Miyadera T et al 2017 *J. Appl. Phys.* **121** 115501

[44]   Shen P, Nie K, Sun X, et al 2016 *physica status solidi (RRL) – Rapid Research Letters* **10** 677